\documentclass[twocolumn,showpacs,showkeys,preprintnumbers,amsmath,amssymb]{revtex4}
 \usepackage{dcolumn}
 \usepackage{bm}
\usepackage{graphicx}
\usepackage{subfigure}
 \begin{document}

 \title{Numerical study of superradiant instability for charged
  stringy black hole-mirror system}

 \author{Ran Li}

 \thanks{Electronic mail: liran.gm.1983@gmail.com}

 \author{Junkun Zhao}

 \affiliation{Department of Physics,
 Henan Normal University, Xinxiang 453007, China}

 \begin{abstract}

 We numerically study the superradiant instability of
 charged massless scalar field in the background of charged stringy black hole
 with mirror-like boundary condition. We compare the numerical result with the
 previous analytical result and show the dependencies of this instability upon
 various parameters of black hole charge $Q$, scalar field charge $q$, and mirror
 radius $r_m$. Especially, we have observed that imaginary part of BQN frequencies grows with the
 scalar field charge $q$ rapidly.

 \end{abstract}



 \maketitle

 Black hole is one of the most fascinating predictions of general relativity.
 The classical black hole described by the solution
 of Einstein field equations is a spacetime region from which gravity prevents anything,
 including light, from escaping. However, energy can be extracted from rotating black holes by impinging bosonic wave with certain frequency condition. This is the well-known classical effect, superradiance \cite{zeldovich,bardeen,misner,starobinsky}.

 Superradiant effect may lead to instability of system combined by rotating black hole
 and bosonic field perturbation. The necessary condition of superradiant instability
 includes two aspects. One is that the black hole should be rotating.
 Under this condition, bosonic wave impinging on such black hole will undergo 
 classical superradiance process. The second is the existence of a potential well outside the horizon
 that can trap the metastable bound states. In some cases, the mass of bosonic
 field or AdS boundary of black hole spacetime may provide such potential well.
 The superradiant instabilities for these cases have been extensively studied in literatures
 \cite{cardoso2004bomb,Rosa,Lee,leejhep,jgrosa,hod2013prd,hodbhb,
 kerrunstable,detweiler,strafuss,dolan,
 Hod,hodPLB2012,konoplyaPLB,DiasPRD2006,zhangw,dolanprd2013,
 cardoso2004ads,cardoso2006prd,KKZ,aliev,uchikata,rlplb,zhang,
 knopolya,rlepjc,clement,randilaton}.

 For a charged scalar field in the background of a charged black hole,
 if the frequency of scalar field satisfies the superradiant condition,
 the wave will also undergo superradiant process \cite{bekenstein}.
 However, it is proved by Hod in \cite{hodrnplb2012,hodrnplb2013} that,
 Reissner-Nordstr\"{o}m (RN) black holes are stable under the perturbations of massive charged scalar fields.

 Soon after, Degollado et. al. \cite{Degolladoprd,Degollado}
 studied the system composed by RN black hole, reflecting mirror, and
 charged scalar field. This is the analogous black hole bomb firstly suggested by
 Press and Teukolsky \cite{press}. The mechanism of black hole bomb seems very simple.
 If one places a reflecting mirror outside of the black hole,
 the wave will be bounced back and forth between the event horizon and the mirror amplifying
 itself each time due to superradiant effect. Meanwhile, the energy of this wave can become sufficiently big in this black hole mirror system until the mirror is destroyed.
 In \cite{Degolladoprd,Degollado}, they found that the instability in the charged case has a shorter
 time scale than in the rotating case. This motivates us to investigate whether the other charged black holes have similar properties as RN black hole.

 In \cite{liprd}, we shown that
 the charged stringy black hole is stable against the massive charged
 scalar perturbation. In \cite{liepjc2014}, we have studied the superradiant instability
 of the scalar field in the background of charged stringy black hole
 due to a mirror-like boundary condition. The analytical expression of the
 unstable superradiant modes is derived by using the asymptotic matching method \cite{cardoso2004bomb}.
 In this paper, we will provide a numerical study of the superradiant instability for charged
 stringy black hole-mirror system, and compare the the numerical results with the
 analytical results.

 This black hole is a the static spherical symmetric charged black holes
 in low energy effective theory
 of heterotic string theory in four dimensions, which is firstly found by
 Gibbons and Maeda in \cite{GM} and independently
 found by Garfinkle, Horowitz, and Strominger in \cite{GHS} a few years later.
 The metric is given by
 \begin{eqnarray}
 ds^2&=&-\left(1-\frac{2M}{r}\right)dt^2+\left(1-\frac{2M}{r}\right)^{-1}
 dr^2\nonumber\\
 &&+r\left(r-\frac{Q^2}{M}\right)(d\theta^2+\sin^2\theta d\phi^2)\;,
 \end{eqnarray}
 and the electric field and the dilaton field
 \begin{eqnarray}
 &&A_t=-\frac{Q}{r}\;,\nonumber\\
 &&e^{2\Phi}=1-\frac{Q^2}{Mr}\;.
 \end{eqnarray}
 The parameters $M$ and $Q$ are the mass and electric charge
 of the black hole respectively.
 The event horizon of black hole is located at $r=2M$.

 In this paper, for simplicity, we only consider the charged massless scalar field
 perturbation in the background of charged stringy black hole. The dynamics of the scalar field
 is then governed by the Klein-Gordon equation
 \begin{eqnarray}
 (\nabla_\nu-iqA_\nu)(\nabla^\nu-iqA^\nu)\Psi=0\;,
 \end{eqnarray}
 where $q$ denotes the charge of the scalar field.
 By taking the ansatz of the scalar field
 $\Psi=e^{-i\omega t}R(r)Y_{lm}(\theta,\phi)$,
 where $\omega$ is the conserved energy of the mode,
 $l$ is the spherical harmonic index, and $m$ is the
 azimuthal harmonic index with $-l\leq m\leq l$,
 one can deduce the radial wave equation in the form of
 \begin{eqnarray}
 \Delta\frac{d}{dr}\left(\Delta \frac{dR}{dr}\right)+UR=0\;,
 \end{eqnarray}
 where we have introduced a new function $\Delta=\left(r-r_+\right)\left(r-r_-\right)$
 with $r_+=2M$ and $r_-=Q^2/M$,
 and the potential function is given by
 \begin{eqnarray}
 U=\left(r-\frac{Q^2}{M}\right)^2(\omega r-qQ)^2-\Delta l(l+1)\;.
 \end{eqnarray}

 The superradiant condition of the charged scalar field is given by \cite{dilatonsr,liprd}
 \begin{eqnarray}
 \omega<q\Phi_H\;,
 \end{eqnarray}
 with $\Phi_H=\frac{Q}{2M}$ being the electric potential at the horizon.
 In this paper, will impose the mirror's boundary condition that the scalar field
 vanishes at the mirror's location $r_m$, i.e.
 \begin{eqnarray}
 \Psi(r=r_m)=0
 \end{eqnarray}
 The complex frequencies satisfying the purely ingoing boundary condition at the black hole
 horizon and the mirror-like boundary condition are called boxed quasinormal (BQN) frequencies \cite{cardoso2004bomb}. In the following, we will present an numerical study of BQN frequencies
 and compare the numerical results with the analytical results.

 With the assume that the Compton wavelength of scalar particle
 is much larger than the typical size of black hole, i.e.
 $1/\omega\gg M$, we can analytically calculate the BQN frequencies of the system by
 employing matched asymptotic expansion method.  
 The analytical expressions for the BQN frequencies are obtained in \cite{liepjc2014},
 which are given by  
 \begin{eqnarray}
 \omega_{BQN}=\frac{j_{l+1/2,n}}{r_m}+i\delta\;,
 \end{eqnarray}
 where the imaginary part $\delta$ is given by
 \begin{eqnarray}
 \delta=-\gamma \left(\frac{j_{l+1/2,n}}{r_m}-q\Phi_H\right) \frac{(-1)^lJ_{-l-1/2}(j_{l+1/2,n})}{J'_{l+1/2}(j_{l+1/2,n})}\;,
 \end{eqnarray}
 with
 \begin{eqnarray}
 \gamma=\frac{2}{(2l+1)}\left(\frac{l!}{(2l-1)!!}\right)^2
 \frac{r_+(r_+-r_-)^{2l+1}}{r_m(2l)!(2l+1)!}&&\nonumber\\
  \times\left(\prod_{k=1}^{l}(k^2+4\varpi^2)\right)
  \left(\frac{j_{l+1/2,n}}{r_m}\right)^{2l+1}\;.&&
 \end{eqnarray}

 From these expression, it is easy to see that, in the superradiance regime,
 $\textrm{Re}[\omega_{BQN}]-q\Phi_H<0$,
 the imaginary part of the complex BQN frequency
 $\delta>0$. This indicates that the BQN frequencies in the superradiant regime
 is unstable for the charged scalar field with the mirror-like boundary condition
 in the background of the charged stringy black hole.

 The numerical methods employed in this problem are based on the shooting method and numerical
 minimization, which is also called the direct integration (DI) method \cite{Degolladoprd, Dolanprd2010,
 Cardosoprd2014,Uchikata}. The DI method
 is specially suited to find the unstable modes, since these modes have the positive
 imaginary parts and therefor decay exponentially at spatial infinity. We can use the DI method
 to solve the BQN frequencies directly from Eq.(4).
 First, near the horizon $r=2M$, we impose the ingoing boundary condition
 \begin{eqnarray}
 R(r)\sim e^{-i(\omega-q\Phi_H)r^*}\;,
 \end{eqnarray}
 where $r^*$ is the tortoise coordinate defined by $\frac{dr^*}{dr}=\frac{r_+-r_-}{r_+}\frac{r^2}{\Delta}$,
 and expand the radial function $R(r)$ as a generalized power series in $(r-r_+)$.
 Then, we can integrate the radial equation (4) with the ingoing boundary outwards from $r=r_+(1+\epsilon)$ and stop the integration at the radius of the mirror. In this procedure,
 we have taken the small $\epsilon$ as $10^{-6}$. The procedure can be repeated
 by varying the frequency $\omega$ until the mirror boundary condition $R(r_m)=0$
 is reached with the desired precision. We can use a numerical root finder to locate the zeros
 of the boundary condition in the complex$-\omega$ plan. The obtained frequency is just the BQN frequency.

\begin{figure}
\subfigure{\includegraphics{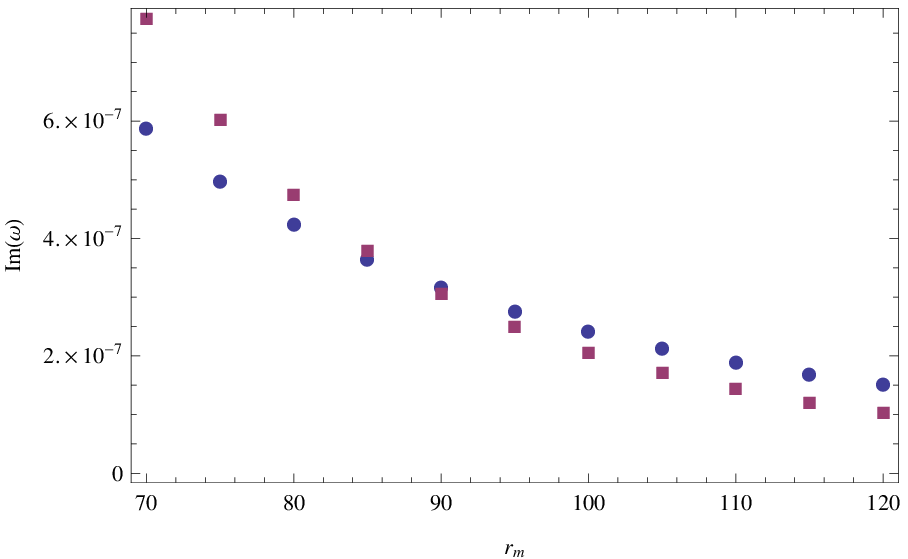}}
\subfigure{\includegraphics{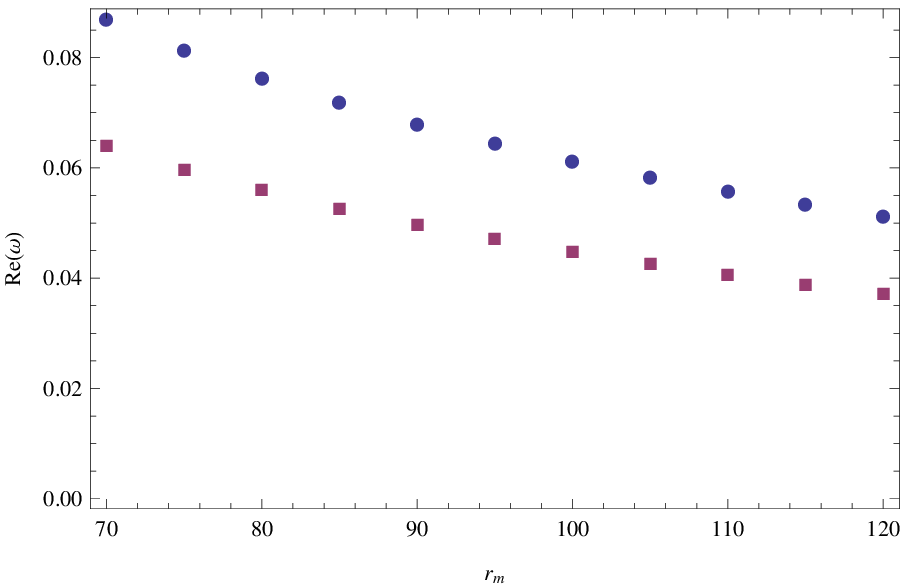}}
\caption{Real and imaginary parts of BQN frequencies for
 $M=1$, $Q=1$, $q=1$ and for different $r_m$. The Bold dots
 and the square dots represent the numerical and the analytical
 results respectively.}
\end{figure}

\begin{figure}
\subfigure{\includegraphics{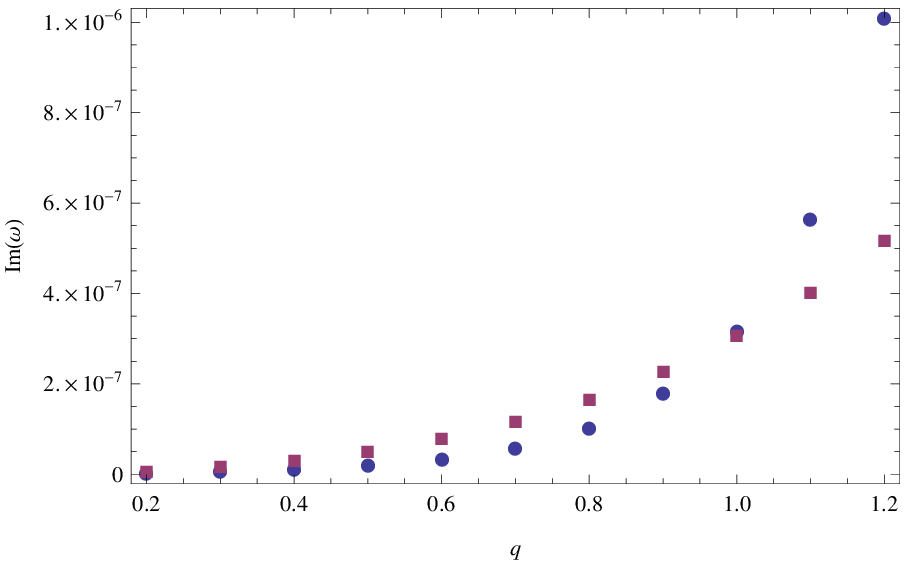}}
\subfigure{\includegraphics{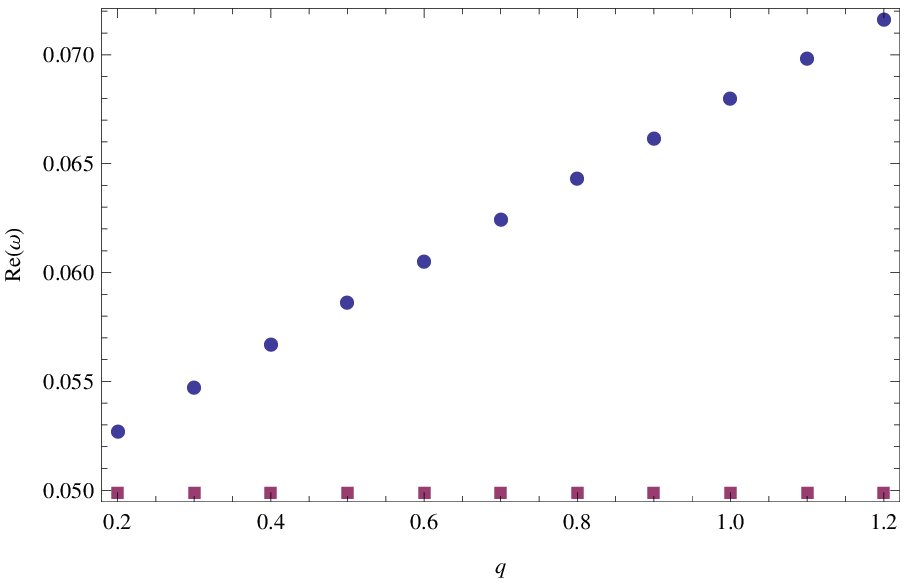}}
\caption{Real and imaginary parts of BQN frequencies for
 $M=1$, $Q=1$, $r_m=90$ and for different $q$. The Bold dots
 and the square dots represent the numerical and the analytical
 results respectively.}
\end{figure}

 Since our interest is to study the superradiant instability
 of this black hole, in which the lower order modes is expected
 to be stronger, we only focus on the $l=1$ mode in the following.
 Firstly, we make a comparison of the numerical and analytical
 results. In Fig. 1 and Fig. 2, we shown the imaginary and the real parts of BQN frequencies
 for different values of the mirror's radius $r_m$ and the charge of scalar field $q$.
 We observe that in these figures the present imaginary part of numerical results are
 in good agreement with the analytical ones, while the real part is not. Especially, from
 Fig. 2, one can see that the numerical results match the analytical very well in the lower frequencies.
 The reason is that the matching technique employed in the analytical calculation
 is expected to yield a better approximation in the lower frequency region.

In Fig. 3, we fix the mass and the charge of black hole, i.e. $M=1$, $Q=1$.
We display the imaginary part
of BQN frequencies as a function of mirror radius $r_m$ for different values of scalar charge $q$.
We observe that when the charge of scalar field decreases, the magnitude of the
imaginary part of BQN frequencies decreases correspondingly. That is to say the instability
is hard to generate for the small charge of scalar field. We also observe that,
for the given values of $M, Q$ and $q$, there exists a critical radius
of the mirror. When the radius is smaller than this critical values, there
is no instability. Clearly, when the charge of scalar field increases,
this critical radius decreases correspondingly.
This point has also been observed in \cite{liepjc2014}
where the analytical calculation was present.
From the analytical result Eq.(9), one can obtain the critical radius can
be approximately given by
\begin{eqnarray}
r_m^{crit}=\frac{j_{l+1/2,n}}{q\Phi_H}\;.
\end{eqnarray}
It should noted that this analytical expression is only valid for the case $qQ\ll 1$.
The numerical results indicate that when the scalar charge $q$ or the black hole $Q$ increases,
the critical radius decreases, which can also be partly observed in Fig.4.
However, the analytical expression Eq.(12) for the critical radius can only 
partially explain the numerical result because Eq.(12) is only valid in the regime $qQ\ll 1$. 
So, there requires an analytical calculation of BQN frequencies for other parameter regime. 
It seems to be difficult to perform such a calculation. In \cite{hod2013prd}, the
analytical calculation of BQN frequencies for the Reissner-Nordstr\"{o}m black hole in the 
regime $qQ\gg 1$ is obtained. It will be interesting to perform an analytical 
treatment of the black hole-mirror system studied in this paper in the $qQ\gg 1$ regime.

\begin{figure}
\subfigure{\includegraphics{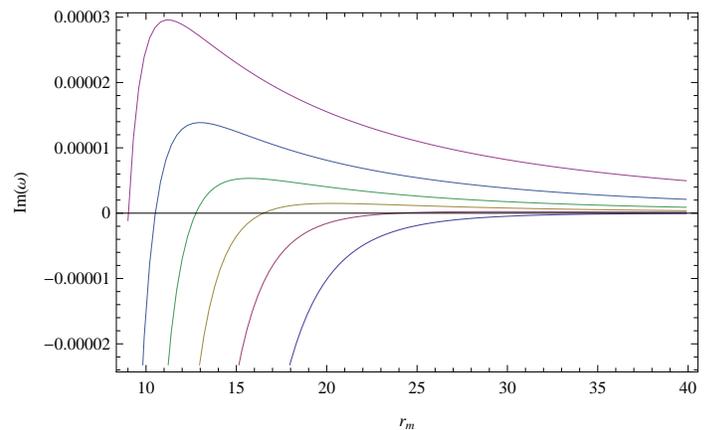}}
\caption{Imaginary parts of BQN frequencies for
 $M=1$, $Q=1$ and for different $q=0.2, 0.4, 0.6, 0.8, 1.0, 1.2$ from bottom up respectively.}
\end{figure}

\begin{figure}
\subfigure{\includegraphics{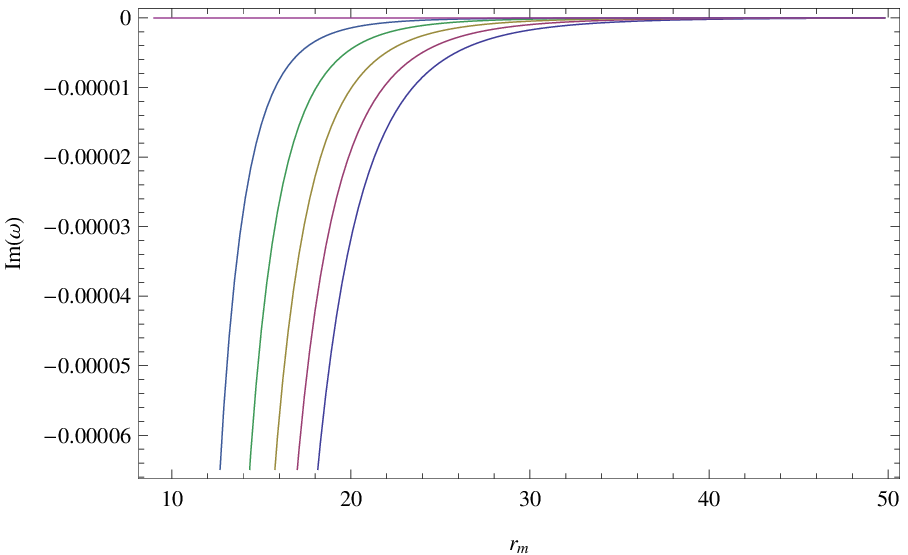}}
\subfigure{\includegraphics{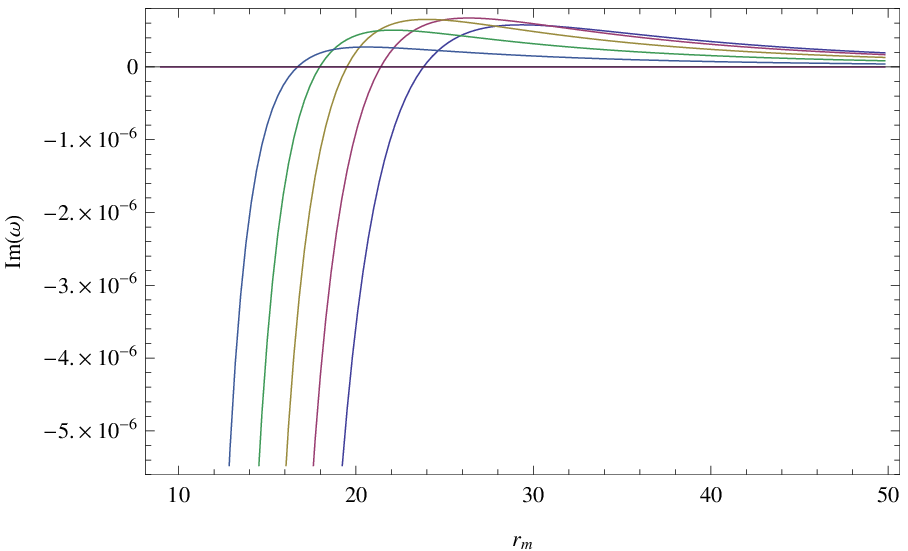}}
\subfigure{\includegraphics{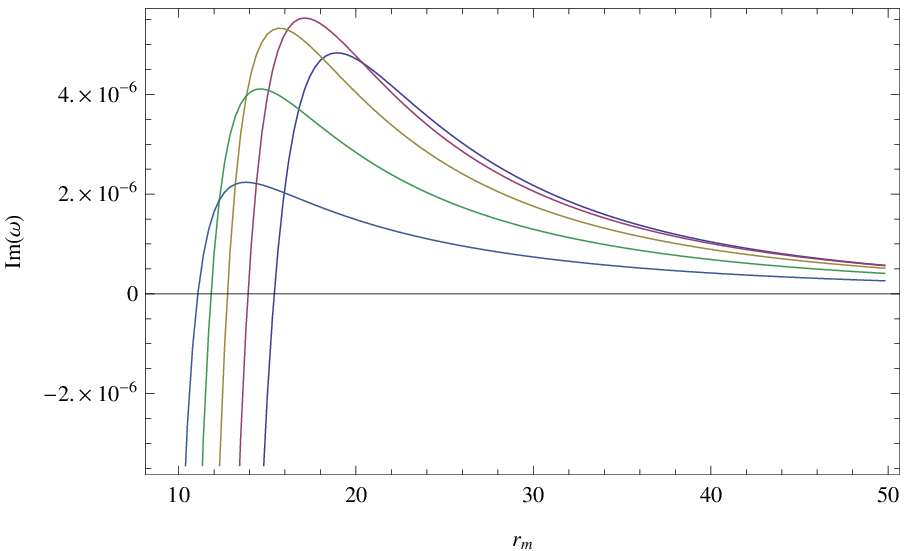}}
\caption{Imaginary parts of BQN frequencies drawn as a function of the mirror radius $r_m$ for
 various values of the black hole charge $Q$ and the scalar charge $q$. We took the black hole
 mass $M=1$. The scalar charge $q$ is equal to $0.2, 0.5, 0.8$ for the first, second, and third
 figure. For each figure, the black hole charge $Q$ is equal to $0.8, 0.9, 1.0, 1.1, 1.2$ from
 right to left.}
\end{figure}

In Fig. 4, we have drawn the imaginary parts of BQN frequencies as a function of the mirror radius $r_m$ for various values of the black hole charge $Q$ and the scalar charge $q$.
From these figures, we firstly observe that, for the small scalar charge $q$, the superradiant
instability is very hard to generate. While for the large scalar charge $q$, it is clear
the instability is easy to generate. This conclusion has been also obtained from Fig.3.
Form the second and third figures, we can also see that when the black hole charge $Q$ increases the critical mirror radius which marks the boundary between the stable and unstable black hole-mirror-scalar field configurations decreases. This verifies the approximate expression Eq.(12) for the
critical radius.

\begin{figure}
\subfigure{\includegraphics{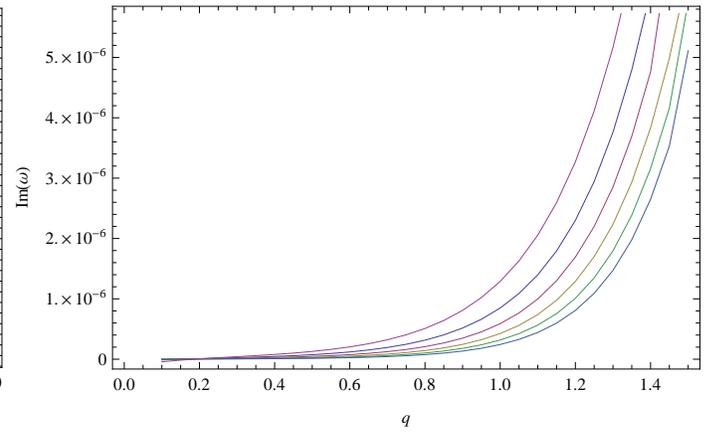}}
\caption{The imaginary parts of BQN frequencies drawn as a function of scalar charge $q$ for $M=1$,
$Q=1$, and various values of mirror radius $r_m$. From top to bottom, $r_m$ is equal to
$50, 60, 70, 80, 90, 100$ respectively.}
\end{figure}

 From the previous numerical conclusion, we expect that the combined 
 system of black hole and scalar field may become extremely unstable
 for large scalar charge $q$ and small mirror radius $r_m$.
 So we have studied the unstable modes
 as a function of scalar charge $q$ and mirror radius $r_m$. In Fig.5,
 we have drawn the imaginary parts of BQN frequencies as a function of scalar charge $q$ for various values of mirror radius $r_m$.
 From Fig. 5, we observe that the imaginary part of BQN frequencies grows with 
 scalar charge $q$ rapidly. This suggests the small growth time scale of the unstable
 superradiant mode for the large scalar field charge. However, the numerical
 calculation of BQN frequencies for the large $q$ becomes difficult which is also mentioned in
 \cite{Degolladoprd}. We have only plotted the results where $q$ lies in the range of
 $0.1$ to $1.5$. As mentioned above, the analytical treatment of BQN frequencies for 
 this black hole-mirror system in the regime of $qQ\gg 1$ may make up for deficiencies in numerical 
 calculations. From Fig. 5, we can also observe that the imaginary part of BQN frequencies grows
 when mirror radius $r_m$ decreases.

 At last, we make a comparison of the numerical results 
 for charged stringy black holes with the numerical results 
 for RN black holes in \cite{Degolladoprd}. 
 In Table I, we show the numerical results of BQN frequencies for
 the RN black hole and the charged stringy black hole. 
 The numerical results of BQN frequencies for
 the RN black hole is adopted from the Table I 
 in Ref.\cite{Degolladoprd}. 
 It is clearly that the RN black hole-mirror system 
 is more unstable than the charged stringy black hole-mirror system. 

 \begin{table}[tbp]
\centering  
\begin{tabular}{lccc}  
\hline
$q$ & $\omega_{RN}$ & $\omega_{CS}$ \\ \hline 
$1.2$\;\; &  $0.0605+7.1405\times10^{-7}i$\;\;   &$0.0606208 + 3.89657\times 10^{-7}i$ \\
$1.6$\;\; &  $0.0657+3.8595\times 10^{-6}i$\;\;  &$0.0658496 + 2.21487\times 10^{-6}i$ \\
 \hline
\end{tabular}
\caption{ The numerical results of RN black hole and 
 charged stringy black hole for $r_m=100$ and $Q=0.8$.}
\end{table}

 In summary, using the direct integration method,
 we have numerically studied the superradiant instability of 
 charged massless scalar field with mirror-like boundary condition
 in the background of charged stringy black hole.
 By comparing the numerical results with the analytical approximation,
 we conclude the analytical calculation is only efficient in the low frequency region.
 We also show the dependencies of this instability upon
 the various parameters of black hole charge $Q$, scalar field charge $q$, and the mirror
 radius $r_m$. Especially, we have observed that the imaginary part of BQN frequencies grows with the
 scalar charge rapidly. We have also compare the numerical results
 for charged stringy black holes with the numerical results
 for charged Reissner-Nordst\"{o}m black holes in \cite{Degolladoprd}, and
 found that the Reissner-Nordst\"{o}m black hole-mirror system is more 
 unstable. At last, we should point out that the analytical
 treatment of BQN frequencies for the charged stingy black hole-mirror system 
 in the regime of $qQ\gg 1$ is required in the future.

 \section*{ACKNOWLEDGEMENT}

 The authors would like to thank Dr. Heng Guo and Hongbao Zhang
 for useful discussion on numerical method. The author Ran Li would also like to thank
 the hospitality at KITPC (Beijing), where this paper was partially completed.
 This work was supported by NSFC, China (Grant No. 11205048).

 \end{document}